\documentclass[twocolumn]{svjour3}
\usepackage[utf8]{inputenc}
\usepackage{hyperref}

\usepackage[dvipsnames]{xcolor}
\usepackage{graphicx}

\usepackage{enumerate}

\usepackage[numbers]{natbib}

\usepackage[framemethod=TikZ]{mdframed}

\usepackage{subcaption}
\usepackage{soul}
\usepackage[normalem]{ulem}

\usepackage{epsf,picinpar}
\usepackage{varioref}
\usepackage{fdsymbol}

\usepackage{pifont}
\usepackage{fancyhdr}

\usepackage{centernot}
\usepackage[makeroom]{cancel}

\usepackage{footmisc}
\usepackage{fp}
\usepackage{xspace}
\usepackage[pscoord]{eso-pic}

\definecolor{linkblue}{RGB}{4,0,237}

\newcommand{\secref}[1]{Section~\ref{#1}}

\newcommand{\figref}[1]{Figure~\ref{#1}}

\newcommand{\hreffinternal}[3]{\href{#1}{\textcolor{#3}{#2}}}

\newcommand{\hreff}[2]{\hreffinternal{#1}{#2}{linkblue}}

\newcommand*{\affaddr}[1]{#1} 
\newcommand*{\affmark}[1][*]{\textsuperscript{#1}}

\mdfsetup{%
   backgroundcolor=gray!3,
   roundcorner=3pt%
}

\newcommand{\placetextbox}[3]{
  \setbox0=\hbox{#3}
  \AddToShipoutPictureFG*{
    \put(\LenToUnit{#1\paperwidth},\LenToUnit{#2\paperheight}){\vtop{{\null}\makebox[0pt][c]{#3}}}%
  }%
}%

\smartqed  
\journalname{Software and System Modeling}

\begin{document}

\placetextbox{0.5}{0.99}{\large\colorbox{gray!10}{\textcolor{WildStrawberry}{\textbf{Author pre-print.}}}}%

\placetextbox{0.5}{0.97}{\large\colorbox{gray!10}{\textcolor{WildStrawberry}{\textbf{The final publication appeared in the Journal of Software and Systems Modeling (SoSyM).}}}}%

\placetextbox{0.5}{0.05}{\colorbox{gray!10}{\textcolor{WildStrawberry}{\textbf{Author pre-print. Final publication available at} \hreff{https://doi.org/10.1007/s10270-024-01154-4}{https://doi.org/10.1007/s10270-024-01154-4}.}}}%

\title{Circular Systems Engineering}

\author{
    Istvan David\protect\affmark[1] \and
    Dominik Bork\protect\affmark[2] \and
    Gerti Kappel\protect\affmark[2]
}

\authorrunning{I. David et al.}

\institute{
    \affaddr{\affmark[1]Department of Computing and Software, McMaster University, Canada}\\
    \affaddr{\affmark[2]Business Informatics Group, TU Wien, Austria}\\
}

\date{Received: date / Accepted: date}

\maketitle

\begin{abstract}
    The perception of the value and propriety of modern engineered systems is changing. In addition to their functional and extra-functional properties, nowadays' systems are also evaluated by their sustainability properties.
    The next generation of systems will be characterized by an overall elevated sustainability---including their post-life, driven by efficient value retention mechanisms. Current systems engineering practices fall short of supporting these ambitions and need to be revised appropriately.
    In this paper, we introduce the concept of circular systems engineering, a novel paradigm for systems sustainability, and define two principles to successfully implement it: end-to-end sustainability and bipartite sustainability.
    We outline typical organizational evolution patterns that lead to the implementation and adoption of circularity principles, and outline key challenges and research opportunities.
\end{abstract}


\section{Introduction}

The steadily accelerating innovation pathways of humankind have rendered prevailing systems engineering paradigms unsustainable. By Brundtland’s classic definition of sustainability~\citep{brundtland1987our}, systems engineering falls short of ``\textit{meeting the needs of the present without compromising the ability of future generations to meet their own needs}''.
By the terms of the four essential sustainability dimensions of \citet{penzenstadler2013generic}, prevalent systems engineering practices fail to fulfill important technical (long-term usage), economic (financial viability), environmental (reduced impact), and social (elevated utility) sustainability principles.

The environmental impacts of systems engineering are particularly apparent.
Manufacturing industries account for nearly a third of the global energy consumption and produce 36\% of CO$_2$ emissions worldwide~\cite{meng2019milp}.
Important end-user systems, such as automobiles, remain significant sources of pollution despite the intensive electrification efforts~\cite{williams2023autogeddon}, and evidence suggests that officially reported laboratory-projected CO$_2$ values might not reflect actual performance~\cite{fontaras2017fuel}.
Information and Communications Technology (ICT), which systems engineering heavily relies on, currently contributes to about 2-4\% of global CO$_2$ emissions, comparable to the carbon emissions of the avionics sector. Without intervention, this number is projected to increase to about 14\% by 2040~\cite{belkhir2018assessing}. However, to follow suit with the rest of the economy, the ICT sector should---directly or indirectly---decrease its CO$_2$ emissions by 42\% by 2030, 72\% by 2040, and 91\% by 2050~\cite{ituict}. The hardware aspects of systems engineering do not look promising either. The amount of discarded electrical or electronic devices, known as e-waste, is recognized by the World Economic Forum as the fastest-growing category of waste~\cite{wef2019new}.
This environmental pressure, coupled with the lack of pronounced efforts for long-term usage of systems through evolution, reuse, and repurposing, paints a particularly daunting picture of the sustainability of systems and engineering methods.

Clearly, systems engineering must change course.
While most experts agree that effective change will require major political and industry intervention~\cite{freitag2021real}, and some argue that fulfilling these goals is not feasible due to the limitations of our current frame of thinking~\cite{becker2023insolvent},
it is also expected that, in the decade ahead of us, users and organizations will reward and demand efforts toward sustainability.
The importance of reuse and repurposing will increase and become a key driving force in systems engineering. Sustainability as a characteristic will become a leading principle in the design, operation, maintenance, and post-life of systems. These trends have been collectively identified in the Systems Engineering Vision 2035 report~\cite{incose2022vision} of the International Council on Systems Engineering (INCOSE) as the top ``\textit{global megatrend}'' that will instigate the development of radically new systems engineering frameworks, methods, and tools. The European Commission has also identified sustainability as a critical enabler of a more resilient European industry within the framework of Industry 5.0~\cite{ec2020industry}.
Expert voices call for immediate action in devising frameworks, methods, and tools for sustainable systems engineering practices~\cite{vanderaalst2023sustainable} and fostering a circular economy~\cite{kristoffersen2020smart}.

Digital transformation trends chiefly associated with Industry 5.0 have created opportunities---such as highly evolved digital capabilities, access to large volumes of data, and a better view of the end-to-end engineering endeavor---to introduce sustainability into systems engineering~\cite{machado2020sustainable}. To leverage these opportunities, we introduce \textit{circular systems engineering}, a paradigm that adopts the circularity principle of circular economy~\cite{corona2019towards} for systems engineering, and by that, enables truly sustainable systems engineering.

The rest of this article is structured as follows.
In \secref{sec:background}, we provide an overview of the background related to our work.
In \secref{sec:circularse}, we offer a definition of circular systems engineering.
In \secref{sec:principles}, we discuss the foundational principles of circular systems engineering.
In \secref{sec:levels-strategies}, we discuss sustainability maturity levels and strategies to reach the level of circularity in systems engineering.
In \secref{sec:challenges}, we outline the key challenges and research opportunities for the modeling community in implementing circular systems engineering.
In \secref{sec:conclusion}, we draw the conclusions.
\section{Background}\label{sec:background}

In this section, we give a brief overview of the background of our work.

\subsection{Sustainability}

The most commonly used view of sustainability originates from Brundland~\cite{brundtland1987our} who defines sustainability as the capacity to ``\textit{meet the needs of the present without compromising the ability of future generations to meet their own needs}''. Brundland differentiates between three aspects of sustainability: economic (financial viability), environmental (reduced ecological impact, e.g., waste), and societal (elevated utility for society and the human). In an effort to adopt sustainability principles for software-intensive and technological systems, \citet{penzenstadler2013generic} extend these aspects with a fourth one: technical sustainability, which describes the ability of a system to be used over a prolonged period. A similar notion of sustainability is voiced by \citet{hilty2006relevance} who define sustainability as the capacity to ``\textit{preserve the function of a system over an extended period of time}''.
We consider the definition of the four-dimensional model of \citet{lago2015framing} (economic, environmental, social, technical) sufficient for our purposes and for the remainder of this article.

Different dimensions of sustainability require different efforts.
Technical sustainability of software-intensive systems is achieved through evolution mechanisms~\cite{durdik2012sustainability,david2023towards}, typically approached at the architectural level~\cite{venters2018software}.
Environmental sustainability is often associated with resource recreation and pollution management. For example, Daly's three principles of achieving sustainability~\cite{daly1990some} demand renewable resources to be used no faster than the rate at which they regenerate; non-renewable resources to be used no faster than renewable substitutes for them can be put into place; and pollution and waste to be emitted no faster than natural systems can absorb them, recycle them, or render them harmless.
\subsection{Circular economy}\label{sec:ce}
Circular economy (CE) is a model of production and consumption in an economic system that promotes retaining value over numerous cycles of the system as long as possible. CE is one among several solutions for fostering sustainable systems~\cite{corona2019towards,garetti2012sustainable}.
The CE concept has been around for decades, mostly attributed to \citet{pearce1989economics}.
Following a comprehensive study on its exact definition, \citet{geissdoerfer2017circular} define CE as ``\textit{a regenerative system in which resource input and waste, emission, and energy leakage are minimized by slowing, closing, and narrowing material and energy loops. This can be achieved through long-lasting design, maintenance, repair, reuse, remanufacturing, refurbishing, and recycling}''.
An important momentum of CE is that it is restorative or regenerative by intention and design~\cite{emf2013towards}. In economies at scale, this typically necessitates implementing proper policies and legal frameworks that promote circularity, such as the Industry 5.0 framework of the European Commission.

Typical forms of value retention in CE are the R-imperatives, such as repairing, reusing, refurbishing, and recycling existing systems, products, and materials. R-frame\-works collect, curate, and organize these R-imperatives. One of the most known of such R-frame\-works is the \textit{3R} which advocates for reducing consumption, reusing products, and eventually recycling materials.
Different domains adopt their own finer or coarser-grained R-imperatives with varying rigor and details. A thorough cross-domain analysis has been provided by \citet{reike2018circular}.

Van der Aalst et al.~\cite{vanderaalst2023sustainable} find that the 10R framework of \citet{reike2018circular} is particularly well-suited to support sustainable systems engineering. Although with the additional refined R-imperatives 10R indeed provides a proper coverage of systems engineering concerns, the activities of R-frameworks are generally too high-level to be actionable, and lack mechanisms for the formal assessment of sustainability under these strategies.
A substantial amount of research has been done on reuse in the realm of digitized engineering domains, such as software~\cite{krueger1992software}. Nowadays, it is hard to encounter top software conferences without dedicated tracks or focus topics on reuse, repurposing, or other kinds of value retention. As a consequence, the benefits~\cite{mikkonen2019software} and threats~\cite{gkortzis2021software} of reuse are fairly understood in the software engineering community. The rich literature on the topic merits a deeper look from a systems engineering point of view to identify methods and techniques that can be \textit{reused} to foster circular systems engineering.
\section{Circular Systems Engineering}\label{sec:circularse}

As put forward by \citet{vanderaalst2023sustainable}, true sustainability requires circularity mechanisms.
The idea of circularity has been around for decades and is generally recognized as a desirable direction for humankind's sustainability and innovation endeavors~\citep{winans2017history}.
While the state of the art considers circular economy~\cite{reike2018circular} as an enabler of sustainability, we argue that promoting circularity to a first principle in systems engineering fosters stronger alignment of technical and sustainability properties of engineered systems.

Adopting the definition of \textit{systems engineering} of INCOSE, the International Council on Systems Engineering~\cite{incose2022systemsengineering} and the definition of \textit{circular economy} of \citet{geissdoerfer2017circular}, we define circular systems engineering as follows.

\begin{mdframed}
Circular systems engineering is the paradigm of designing, developing, operating, maintaining, and retiring systems through sustainable systems principles that foster value retention over multiple engineering cycles.
\end{mdframed}

\textit{Systems principles} range from high-level mental models to well-defined ones, such as causal loop diagrams; include key mechanisms, such as abstraction, modularity, and encapsulation; and integrate into techniques such as systems thinking and systems dynamics. A detailed account of systems principles is given by the Systems Engineering Body of Knowledge (SEBoK)~\cite{adcock2020principles}.
\subsection{Circularity in systems engineering}\label{sec:circularity}

In the definition, \textit{value retention} refers to the act of keeping value as long as possible. Value retention is implemented through the mechanism of \textit{circularity}. Circularity ensures physical and virtual value is reused, renewed, and regenerated, rather than being wasted. Waste is produced when value cannot be circulated anymore~\cite{morseletto2020targets}. While the governing frame of thinking contextualizes sustainability within the lifecycle of one specific system, circular systems engineering emancipates sustainability from these confines and unlocks a more holistic view of sustainability.

We distinguish between two classes of value in systems engineering:
\begin{itemize}
    \item \textbf{physical value}, such as raw materials, energy, components of subsystems; and
    \item \textbf{digital or virtual value}, such as software systems, engineering documents, and the knowledge encoded in trained AI models, simulation traces, and results of experiments.
\end{itemize}

The latter class is typically not explicit in circular economy models (\secref{sec:ce}). However, since software systems, models, and experimental traces are first-class citizens in systems engineering, they must be taken into account by successful implementations of circular systems engineering. For example, retaining the value of a software architecture through evolution embodies technical sustainability~\cite{lago2015framing}; and retaining the value of a costly wind tunnel experiment embodies economic and environmental sustainability.

\paragraph{Retaining physical value.}

Some extended R-frameworks (see \secref{sec:ce}) refine the reuse-recycle mechanisms into more tangible ones. For example, half of the R-imperatives of the  10R framework~\cite{reike2018circular} focus on retaining value by disassembly and reassembly---either with the same components, configuration, and purpose, or with completely new ones.
Component traceability has been a topic of interest in Internet of Production (IoP) initiatives~\cite{jeschke2017industrial,pennekamp2019towards}. In electrical and electronic equipment waste refurbishing services, for example, traceability through transportation is an important enabler of the offering~\cite{sharpe2018cyber} as identifying transported products provides assurance of location, condition, and integrity.
These early results should be further extended to the traceability of components along \textit{multiple} system lifetimes to contribute to more effective circularity mechanisms.
Eventually, system components are retired and disassembled, and in most cases, scrap raw material and energy are the only value that can be salvaged. Understanding material trade-offs, especially using better-choice materials---i.e., functional materials with reduced environmental impact by enhanced durability and lifetime~\cite{shen2023computational}---is paramount for reuse and repurposing. To assist such an understanding, machine learning and AI methods have been proposed, e.g., to accelerate material development by physics-constrained AI~\cite{gomes2019crystal} and design space exploration by active transfer learning and data augmentation~\cite{kim2021deep}. Such advanced mechanisms improve value retention in circular systems engineering, and therefore, their development offers research avenues with elevated utility.

\paragraph{Retaining digital and virtual value.}
Engineering knowledge encoded in digital and virtual value represents a particularly important class of values that should be retained across numerous system lifetimes.
Yet, knowledge retention is not well-understood and is mostly ignored in current circularity strategies and frameworks \cite{corona2019towards,iacovidou2021systems}.
Efficient reuse of virtual value contributes to each aspect of sustainability as it allows for better design both in technical and economic terms, while also allowing for optimized operation and governance~\cite{proper2021towards,poels2022dt4gitm}.
Reusing design knowledge and transposing previously learned lessons has been a topic of particular interest.
Traditional techniques for knowledge retention through the use of ontologies provide rich expressiveness but do not scale to the level of truly complex systems and their engineering processes~\cite{osman2021ontology}. This is partly due to the lack of methods for identifying truly valuable knowledge to be retained. Assessing the value of knowledge is particularly challenging due to its non-monetary, highly abstract, and tacit nature. This is aptly exemplified by the discipline of infonomics~\cite{laney2017infonomics}, which takes a middle ground and provides actionable metrics for information instead of knowledge.
More automated techniques of retaining codified knowledge have been a subject of active research in machine learning. Transfer learning~\cite{weiss2016survey} is the technique of applying previously learned knowledge in congruent tasks. Such techniques have been successfully applied in an array of complex problems, such as atmospheric dust aerosol particle classification~\cite{ma2015transfer} and poverty mapping~\cite{xie2016transfer}, and very much technical problems, such as image recognition~\cite{shao2015transfer}.
\subsection{Impacting systems and methods through leverage points}\label{sec:leveragepoints}

Adopting circularity mechanisms into systems engineering affects both engineered systems and engineering methods. Leverage points of complex systems are places where a small shift in one aspect can produce big changes \cite{meadows1999leverage}. Following the clusters of leverage points by \citet{penzenstadler2018software}, circular systems engineering aims to change the intent of the system and its stakeholders. This covers the following leverage points.
\begin{description}
    \item [\textbf{L3:}] \textbf{Changing the goals of the systems engineering endeavor.} Instead of optimizing for short-term, e.g., for the functional properties of one specific engineered system, operational excellence of organizations in circular systems engineering considers every engineered system, and the goal of the organization becomes the overall sustainability of engineered systems and engineering methods.
    \item [\textbf{L2:}] \textbf{Changing the mindset of stakeholders.} Instead of baking sustainability goals into monetary KPIs, stakeholders in circular systems engineering embrace sustainability as a first principle in systems engineering.
    \item [\textbf{L1:}] \textbf{The power to transcend paradigms.} Instead of sticking to a limited number of methods, systems engineers in circular systems engineering fully adopt a multi-paradigm mindset, such as multi-paradigm modeling~\cite{vangheluwe2002introduction}, and feel empowered to redefine para\-digms for more sustainable alternatives, such as sustainable software engineering~\cite{mcguire2023sustainability}, sustainable digital twinning \cite{bellis2022challenges,heithoff2023digital}, and green AI~\cite{verdecchia2023systematic}.
\end{description}

Through these leverage points, circular systems engineering improves the sustainability outlooks of the systems we engineer, and the methods by which we engineer our systems.

Discussions with technology transfer experts indicate that circular systems engineering might be an important development for industry domains, such as industrial automation~\cite{bangemann2014state}--—the computer-aided technique of alleviating the human bottleneck in advanced industry processes. Driven by stakeholder demands for operational excellence~\cite{dahlgaard1999integrating} and better control over quality and time-to-market, industrial automation is a global market experiencing rapid growth, and could be impacted at any of its L1---L3 leverage points by a circular systems engineering mindset.
\section{Principles of Circular Systems Engineering}\label{sec:principles}

Circular systems engineering relies on two principles.
First, fostering circularity mechanisms requires understanding how and when value is created along the systems engineering process. This is the principle of end-to-end sustainability (\secref{sec:principles-e2e}).
Second, along with the sustainability of the engineered system, the sustainability of the employed engineering methods is equally important. This is the principle of bipartite sustainability (\secref{sec:principles-bipartite}).
Implementing circular systems engineering requires implementing both end-to-end and bipartite sustainability.

\begin{figure*}[htb]
    \centering
    \includegraphics[width=\linewidth]{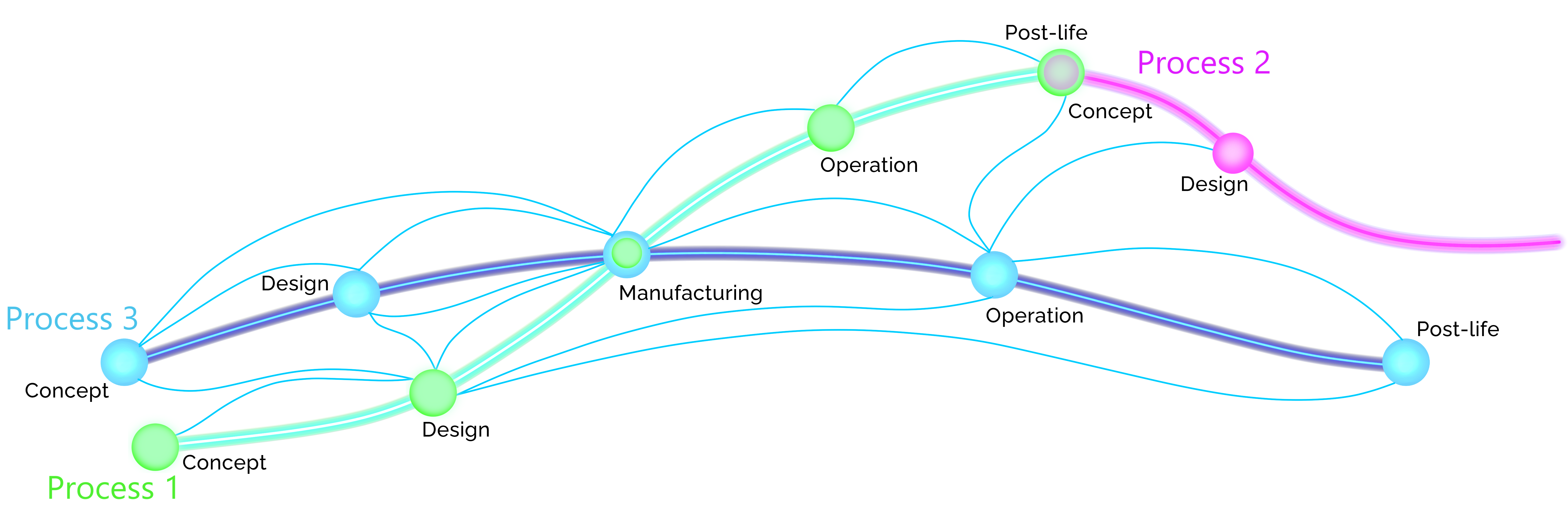}
    \caption{Process network composed of three engineering processes, in which information can freely flow between activities}
    \label{fig:fabric}
\end{figure*}

\subsection{End-to-end sustainability}\label{sec:principles-e2e}

To achieve circularity in systems engineering, a detailed end-to-end understanding of the engineering endeavor is required. That is, reasoning about sustainability must scale from the scope of single engineering activities to the overall engineering process and, eventually, to networks of processes~\cite{oberdorf2023predictive}.
In each scope, reasoning must entail $(i)$ assessment of sustainability and $(ii)$ finding optimal trade-offs among functional and sustainability properties.

\paragraph{Reasoning in the scope of activities.}
Assessing and optimizing for sustainability starts in the scope of single lifecycle phases or activities of the systems engineering process. This scope is typically implied by organizationally and logically isolated engineering silos. Engineering silos come in various sizes and complexity. For example, an engineering silo can be a team of hydraulics experts in the development process of a complex cyber-physical system, or a separate business unit or company integrated into the supply chain.
Typical examples in this scope include reasoning for sustainability at design time~\cite{fatima2023review}, reasoning for sustainability during manufacturing~\cite{henaohernandez2019control}, and reasoning for sustainability at operation time~\cite{franciosi2018maintenance}.
Unfortunately, siloed methods no longer scale with the pace of innovation and the expectations of quality and efficiency~\cite{dertien2021state}. Despite the high degree of digitalization, reasoning about sustainability within engineering silos has limitations due to multiple sources of truth and data inaccessibility, and challenges organizations by performance issues due to duplicated efforts.

\paragraph{Reasoning in the scope of a process.}
Scaling up reasoning to the scope of a single engineering process allows for assessing sustainability and finding trade-offs in the temporal dimension. In this scope, trade-offs might be found between engineering activities along the process.
Reasoning in this scope aligns well with current trends in industrial practice focusing on the digital thread. Singh~\cite{singh2018engineering} defines the digital thread as ``\textit{a data-driven architecture that links together information generated from across the product lifecycle}''. West and Pyster~\cite{west2015untangling} relate the digital thread to traditional Model-based Systems Engineering (MBSE) and its artifacts by defining the digital thread as ``\textit{a framework for merging the conceptual models of the system (the traditional focus of MBSE) with the discipline-specific engineering models of various system elements}''. The digital thread fosters the propagation of streams of information across silos, effectively integrating them along the system lifecycle.
A typical example of finding sustainability trade-offs in this scope might be choosing a less sustainable manufacturing method for more sustainable operation and maintenance characteristics. Analogous ideas have been explored, e.g., in product-assembly co-design~\cite{vanacker2021knowledge}, where the components of the engineered system are optimized for easier integration (assembly).
Despite the improved scope, reasoning about sustainability within a process still leaves room for improvement. Specifically, finding trade-offs across parallel and subsequent engineering processes requires extending the scope of investigation even further.

\paragraph{Reasoning in the scope of process networks.}

Advancements in digitalization and widespread digital transformation triggered organizations to focus more on their business processes and foster end-to-end business process execution and management~\cite{eversheim2013prozessorientierte}. Modern organizations connect departments through enterprise process networks~\cite{page1999observations}, linking departments, processes, and information systems and establishing inter-department and inter-process dependencies.
Investigating sustainability in the scope of process networks allows for assessing sustainability and finding trade-offs by considering inter-process phenomena and properties.
\figref{fig:fabric} shows two typical cases in which parallel and sequential processes overlap and meet each other.
One of the typical activities this scope enables is the rationalization of manufacturing. In \figref{fig:fabric}, the \textit{Manufacturing} phases of \textit{Process 1} and \textit{Process 3} overlap. Synchronizing the manufacturing phases of engineering processes and optimizing production scheduling are major energy-saving factors~\cite{gao2019review} as about 80\% of the energy consumed by machine tools is reportedly attributed to idle state operation~\cite{mouzon2007operational}. Optimizing the manufacturing schedule can substantially improve the energy footprint of organizations. This can be achieved by finding proper schedules across engineering processes. Another typical activity in this scope is design for reuse, i.e., allowing future generations to retain value from a retired system more efficiently. As the post-life phase of one engineering process is channeled into the concept design phase of a subsequent process, substantial value can be retained in the form of raw material, energy, and engineering knowledge.
The scope of process networks is where circularity emerges as value is circulated from one engineered system to the other. This allows for a higher level of sustainability, specifically in environmental and economic aspects. Fostering circular systems engineering, thus, requires an end-to-end understanding of process networks.
\subsection{Bipartite sustainability}\label{sec:principles-bipartite}

True sustainability can only be achieved if sustainable systems are engineered through sustainable methods. We refer to this principle as \textit{bipartite sustainability}. Bipartite sustainability emphasizes that sustainable systems and sustainable engineering methods are equally important and form an inseparable unit when assessing sustainability, as incurring sustainability debts along the engineering process might render the overall, end-to-end systems engineering endeavor unsustainable.

Prime examples of engineering methods that might quickly become unsustainable are machine learning (ML) and artificial intelligence (AI). Given the immense complexity of reasoning about sustainability, modern ML and AI methods will inevitably make their way into the toolbox of circular systems engineering.
For example, developing simulators using various forms of machine learning~\cite{david2022devs,legaard2023constructing} allows for replacing parts of complex control algorithms with AI components, resulting in more efficient calculations.
Such trends have been identified in numerous domains, e.g., affordable and sustainable energy~\cite{UN2015sdg7}, AI for climate change~\cite{rolnick2022tackling}, and AI4Good~\cite{ai4good}.
The abundance of data along digital thread-based engineering practices makes such directions feasible.
However, ML and AI techniques have demonstrated problems in terms of environmental~\cite{verdecchia2023systematic} and social sustainability~\cite{ferrer2021bias}. Training ML/AI models is a particularly resource-demanding endeavor, especially regarding energy. The experiments by \citet{strubell2019energy} report a staggering 270 tonnes of CO$_2$ emission for training a large NLP model with neural architecture search~\cite{so2019evolved}, equivalent to the lifetime emissions of five cars. Google's AlphaGo Zero generated 96 tonnes of CO$_2$ over 40 days of research training~\cite{inhabitat2020mit}, equivalent to the lifetime emissions of two to three cars \citep[Table 1]{strubell2019energy}.
Clearly, to support the sustainability ambitions of circular systems engineering, AI itself must become sustainable. This will require researching and prioritizing computationally
efficient hardware and algorithms~\cite{strubell2019energy}, better tools to assess the environmental impact of AI~\cite{lacoste2019quantifying,henderson2020towards}, and proper legal, societal, and technical frameworks to govern and enforce sustainable AI practices~\cite{vanwynsberghe2021sustainable,coeckelbergh2021ai}.

Another supporting method of sustainability ambitions that itself requires sustainability considerations is digital twinning~\citep{kritzinger2018digital}.
Capgemini reports that 60\% of organizations believe twin technology is critical to improving sustainability efforts~\cite{capgemini2022digital}. Some of the sectors with visible benefits include manufacturing, where multi-million dollar savings are being regularly reported thanks to digital twinning~\cite{wsj2019unilever}; smart cities, where Accenture estimates energy consumption rationalization by 30\%-80\% through digital technologies~\cite{accenture2022critical}; and traditionally lower-digitized sectors, such as biophysical systems and agriculture, where digitalization improves crop-to-energy ratio and substantially stabilizes the supply chain \cite{david2023digital}.
While research primarly focuses on sustainability \textit{by} digital twins, it is important to acknowledge that digital twins themselves need to be sustainable to support larger sustainability goals. \citet{bellis2022challenges} report four sustainability challenges of digital twinning: energy consumption, modeling effort and complexity, the ability to evolve with the physical twin, and deploying the twin architecture within organizations.
The development of digital twins is hindered by the complexity of systems subject to digital twinning. To improve the economic sustainability of digital twins, variability-based techniques have been recommended~\cite{berger2013survey,kang2013variability}, but sound composition mechanisms are still lacking. Closely related work has been done in the synthesis of simulators~\cite{hansen2021synthesizing}, which are key enablers of digital twins. Such preliminary results could be extended to the combined digital-physical realm to support the rapid synthesis of digital twins.
From a technical sustainability point of view, proper evolution mechanisms of digital twins, potentially governed by comprehensive frameworks and taxonomies~\cite{david2023towards}, are an excellent way to move toward more sustainable digital methods. On a related note, self-adaptation has been identified by \citet{fur2023sustainable} as a key mechanism for the technical sustainability of digital twins that require, e.g., flexible architectures, automated re-calibration of simulation models, and ontologies to augment structural information.

Fostering bipartite sustainability requires joint reasoning about the sustainability properties of the engineered system and engineering methods. Similar questions have been the focus of recent emergent techniques, such as product-production co-design~\cite{albers2022product}.

\begin{figure*}[htb]
    \centering
    \includegraphics[width=0.66\linewidth]{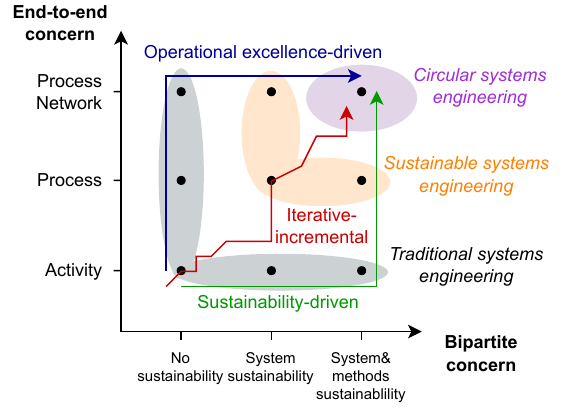}
    \caption{Framework to position sustainability maturity levels and maturation strategies}
    \label{fig:strategies}
\end{figure*} 
\section{Maturity Levels and Strategies Toward Circular Systems Engineering}\label{sec:levels-strategies}

Implementing circular systems engineering requires making strides both in end-to-end and bipartite sustainability. As shown in \figref{fig:strategies}, these two principles span a space of sustainability maturity levels. We now discuss these levels and how organizations can advance through them by improving along the two principles.

\subsection{Sustainability maturity levels}\label{sec:stages}

\paragraph{Traditional systems engineering} typically seeks a balance between the quality, cost, and delivery time of engineered systems. At this level of maturity, efforts primarily aim at a better understanding of these three factors in either of the two dimensions of the framework in \figref{fig:strategies}, but not in combination of the two. This typically leads to advanced process methods in the \textit{end-to-end} dimension of the framework in \figref{fig:strategies}. Substantial work has been dedicated to engineering process modeling~\cite{pahl1996engineering}, assessment~\cite{rout2007spice}, optimization~\cite{david2016engineering}, and adaptation~\cite{seiger2019toward}.
Occasionally, organizations might realize that quality, cost, and time-to-market can be understood through the lens of sustainability, leading to advancement in the \textit{sustainability} dimension of the framework.
For example, sustainability can be interpreted as a property of systems quality~\cite{lago2015framing}, which allows for a sustainability-driven assessment of quality and cost factors. Ensuring technical sustainability~\cite{razavian2014four} also allows for reduced costs by proper evolution after the system has been implemented. Currently, sustainability efforts are mostly focused on energy consumption of software-intensive systems~\cite{ournani2020reducing,palomba2019impact} and advanced digital facilities~\cite{bellis2022challenges}, and on reducing waste of engineering processes~\cite{andersson2011environmental}.

While advancements along the two dimensions might happen, a common trait of traditional systems engineering is the inability to combine advanced process methods with extended sustainability scopes.

\paragraph{Sustainable systems engineering}
combines elements of advanced process methods and extended sustainability scopes. Most importantly, this sustainability maturity level allows reasoning about system-level sustainability along the engineering process and across process networks.
This is demonstrated by sustainability-focused end-to-end methods, such as process scheduling for waste management~\citep{lehesran2019operations}. Another example is manufacturing rationalization~\cite{mouzon2007operational,gao2018neural} enabled by an improved understanding of the end-to-end process network and the resulting improved schedulability. Unfortunately, the stratified nature of sustainability~\cite{mcguire2023sustainability} hinders the development of methods combining process-based longitudinal reasoning with reasoning about sustainability, despite the well-understood benefits~\citep{daoutidis2016sustainability,barisic2023modelling}. Thus, sustainable systems engineering is an ambitious goal for the majority of organizations currently.

A limitation of sustainable systems engineering is its inability to properly reason about value retention loops, i.e., to introduce systematic circularity into the engineering practice. This is due to the lack of ability to combine end-to-end process networks with bipartite systems and method sustainability. This limitation, in turn, motivates the most evolved maturity level of sustainability: circular systems engineering.

\paragraph{Circular Systems Engineering} improves on the final limitation of sustainable systems engineering and allows for combining end-to-end concerns with bipartite sustainability. In this combination of concerns, value retention loops are facilitated through the thorough understanding of $(i)$ the network of systems engineering processes, $(ii)$ the sustainability properties of the engineered system, and $(iii)$ the sustainability properties of the employed engineering methods.
Circular systems engineering enables assessing and optimizing for systems properties, including sustainability properties across multiple engineering endeavors and allows for finding sustainability trade-offs across different engineered systems.
In the most characteristic case, trade-offs can be identified in the upcycling of a systems~\cite{wegener2016upcycling}---and that, at the design phase or lifetime of the original system. In heterogeneous cyber-physical systems, this often might be achieved at a relatively low cost by replacing software that operates physical components. For example, older smartphones can be used as cheap sensors~\cite{li2010smartphone} to manage the sustainability pressure stemming from the constant need to upgrade computing devices due to Moore's law.
Design for reuse~\cite{richardson2011design} and sustainable remanufacturing~\cite{jensen2019creating} are examples of initiatives that can benefit from a circular systems engineering mindset.
Recently, digital twins opened up opportunities for efficient data harvesting and rich decision support based on modeling and simulation~\cite{heithoff2023digital} and might serve as the enablers of this level of maturity.

\subsection{Sustainability maturation strategies}

Maturation strategies from traditional to circular systems engineering can generally follow three pathways.

\paragraph{Operational excellence-driven maturation} aims to achieve a better understanding of the value chain and prioritizes improvements in the end-to-end dimension before dedicating efforts to sustainability.

\citet{dahlgaard1999integrating} define operational excellence (OpEx) by the four `P's of excellence: people, partnerships, processes, and products. Covering the end-to-end value chain is of particular importance in achieving operational excellence. Along an \textit{operational excellence-driven strategy}, a company would develop its internal business capabilities to the level where internal organizations and departments are well-connected through the enterprise process network layer, allowing for the efficient flow of information and advanced analytics techniques such as real-time and predictive analytics~\cite{oberdorf2023predictive} that help achieve excellence.

This strategy might be dominant in sectors where sustainability concerns of engineered systems and, more importantly, engineering methods are dominated by other functional and extra-functional criteria. Such a maturation strategy can be observed in software engineering, where lifecycle models have been researched for decades, but we are just at the beginning of understanding the environmental impacts of computationally demanding methods.

Eventually, an OpEx-driven strategy would implement more elaborate sustainability concerns to arrive at circular systems engineering.

\paragraph{Sustainability-driven maturation} favors improvements in the sustainability dimension and then gradually extends this know-how to the engineering processes and process networks. Focusing exclusively on this dimension (and omitting end-to-end concerns) allows organizations to adopt extensive sustainability management frameworks, such as the ones imposed by ISO 14000~\cite{iso2021environmental} for environmental management, ISO 26000~\cite{iso2222social} for social responsibility, and various other, more domain-specific standards.

This strategy might be dominant in sectors where sustainability concerns are of particular importance, either due to legislation or social expectations. Pertinent examples include industries with increased environmental impact, such as machine production, cyber-physical systems, or the automotive industry.

A sustainability-driven strategy would emancipate sustainability concerns from single engineering activities and impose them on other stages of the systems engineering process in order to arrive at circular systems engineering.

\paragraph{Iterative-incremental maturation} aims to balance the previous two strategies, gradually expanding in end-to-end and sustainability concerns.

Iterative and incremental maturation allows for regular reflections and helps scope subsequent developments better, e.g., by setting the right requirements~\cite{duboc2020requirements}. Evidence also suggests~\cite{lopezalcarria2019systematic} that agile principles, which are inherently iterative and incremental, foster a better understanding of sustainability goals.

This strategy might be the most typical and natural as it allows for quick wins at a particular time and context, and the refined scopes of iteration minimize strategic risks.
\section{Challenges and Research Opportunities}\label{sec:challenges}

System complexity often renders the assessment of sustainability and finding trade-offs among functional and sustainability properties a wicked problem~\cite{rittel1967wicked}. Such wicked problems are often impossible to solve due to the complex and interconnected nature of the components of systems and the ensuing sub-problems. For example, of the seventeen SDGs formulated by the UN, fourteen directly incorporate socio-economic elements (goals 1-12, 16, and 17), rendering their modeling and analysis a particularly challenging problem.
In such a convoluted problem space, modeling, and especially model-based and model-driven engineering techniques~\cite{schmidt2006model}, are powerful tools in taming complexity~\cite{bork2023role}. Through the mechanism of abstraction, model-driven techniques aid understanding of sustainability properties that are often stratified (have different meanings at different levels of abstraction) and multi-systemic (have different meanings for stakeholders of different domains) \cite{mcguire2023sustainability}.

We now outline key challenges and research opportunities for the modeling community in support of achieving higher sustainability maturity levels (\secref{sec:levels-strategies}) in systems engineering.

\subsection{Process methods}\label{sec:challenges-process}

The end-to-end principle of circular systems engineering (\secref{sec:principles-e2e}) necessitates proper modeling and simulation methods for the assessment, optimization, and enactment of systems engineering processes.

\paragraph{End-to-end process modeling.}

Assessing sustainability requires reasoning about the properties of the engineered system in conjunction with the engineering process~\citep{machado2020sustainable,david2018translating}.
The lack of formal rigor in systems engineering process models~\citep{barisic2023modelling} renders the end-to-end assessment and optimization of engineering endeavors an insurmountable challenge.
While isolated efforts have been made, e.g., in sustainable manufacturing~\citep{andersson2011environmental} and business information systems~\citep{mihalewilson2022corporate}, holistic end-to-end approaches for sustainable systems engineering are lacking.
Existing process methods have the potential to support the modeling needs of circular systems engineering. For example, substantial research has been conducted on process methods for finding quality and cost trade-offs in multi-disciplinary design~\citep{david2018translating}. These methods might be capable of supporting technical and economic sustainability aspects of systems engineering. Extensions can be made in terms of supported lifecycle phases, e.g., by addressing the post-life of systems and in terms of externalizing sustainability properties as first-class citizens of complex process models~\cite{challenger2020ftg}. For the latter, environmental sustainability should be considered as an immediate target, e.g., building on activity-based cost modeling and value stream mapping coupled with discrete event simulation~\cite{andersson2011environmental}.
A key challenge to address is the modeling of process networks.
Sound composition mechanisms~\cite{pratt1982composition} could help compose isolated processes into process networks.
Current methods for the comprehensive modeling, analysis, and optimization of such process networks are limited. Notably, process monitoring and prediction frameworks that are able to leverage the entirety of enterprise data sources are completely missing from the state of the art~\cite{oberdorf2023predictive}.

\paragraph{Enactment and control.}
Once engineering processes have been optimized, they must be properly executed. 
However, as highlighted by \citet{daoutidis2016sustainability}, sustainability practices create operational challenges, motivating the employment of well-established process enactment methods.
Process enactment is commonly defined as the use of software to support the execution of operational processes~\cite{lambeau2017process,vanderaalst2003workflow}.
DevOps principles, especially those tailored to digital twins as the primary means of process enactment~\cite{hugues2020twinops} offer particular upside in narrowing the gap between process modeling and process execution. However, research on such advanced lifecycle models is still in a preliminary phase.
Digital process twins~\cite{caesar2020information} further improve enactment and run-time process model adaptation capabilities and require the attention of researchers. A digital process twin is a virtual representation of the real process that captures the process's context and provides control capabilities. In contrast to traditional digital twins of systems or their components, the focus of a digital process twin is the run-time management of the process with typical goals of controlling the quality of the produced product, reducing energy consumption, and ensuring compliance with the process model~\cite{armendia2019evaluation,hanel2019development}.
By that, digital process twins are primary candidates in circular systems engineering to govern and guide day-to-day activities. 
\subsection{Modeling and optimizing for sustainability}\label{sec:challenges-sustainability}

The bipartite sustainability principle of circular systems engineering (\secref{sec:principles-bipartite}) necessitates proper modeling and simulation methods for assessing the sustainability properties of systems and methods and finding trade-offs among these properties.

\paragraph{Sustainability assessment.}
Currently, the evaluation of sustainability in software-intensive systems is mostly a manual effort~\cite{venters2014blind}. There is a pressing need for better automated and more comprehensive methods.
As mentioned at the outset, sustainability is a multi-faceted notion involving many different interrelated aspects. This renders the assessment of sustainability challenging. There are only a few domain-specific languages available for specifying sustainability goals~\cite{gramelsberger2023enabling,lago2015framing} despite the ability of modeling to tame complexity. Given the multi-systemic nature of sustainability, multi-paradigm modeling~\cite{vangheluwe2002introduction}, multi-view modeling~\cite{cicchetti2019multi}, and user-friendly, flexible modeling approaches, such as blended modeling~\cite{david2023blended}, can provide solid foundations for the next generation of sustainability modeling frameworks and tools.
Key challenges in this aspect include the integration of heterogeneous models and the need to personalize specific views and domain-specific concepts for different stakeholders~\cite{combemale2016modeling}. Iterative and incremental evaluation methods~\cite{penzenstadler2021iterative,gramelsberger2023enabling} allow for assessing and re-assessing sustainability properties continuously as systems are developed. However, systematic methods are missing.

Frameworks, such as the Sustainability Awareness Framework~\cite{duboc2019do,brooks2023assessing}
and taxonomies, such as the one by \citet{bischoff2022taxonomy}, help engineers understand the sustainability impacts of systems in all relevant sustainability dimensions. However, even state-of-the-art frameworks fall short of quantifying sustainability and providing actionable insights into sustainability shortcomings.
\citet{tasdemir2018systematic} recommend modeling for sustainability assessment to encompass all verticals down to physical processes, for which Lean methodologies could set the foundation.
Such ideas have been explored in, e.g., sustainable supply chain assessment by value stream mapping~\cite{sparks2014combining} and in environmental cost modeling through activity-based costing~\cite{andersson2011environmental}.

\paragraph{Finding sustainability trade-offs.}
Putting sustainability assessment methods in place enables optimizing systems engineering processes for sustainability goals. Optimization, however, is not trivial given the stratified and multi-systemic nature of sustainability. For example, optimizing towards one SDG target may have negative side-effects on another one~\cite{renaud2022sdgtradeoffs, tzachor2022potential}. The general challenge, therefore, is finding acceptable and responsible balancing trade-offs between different sustainability goals. Adopting the mindset of circular systems engineering allows for further opportunities to find trade-offs across different phases of engineering processes.
As put more explicitly by~\citet{tzachor2022potential}, ``\emph{in terms of the SDGs, trade-offs are guaranteed to arise given the inherent contradictions across the 169 targets}'', and by~\citet{zhang2013sustainablemanufacturing}
``\emph{industry is confronted with the challenge of balancing economic and financial priorities against environmental and social responsibilities}''.
The ability to identify, analyze, and optimize the sustainability trade-offs within and across digital threads by different SDG targets, Key Performance Indicators (KPIs), and Key Environmental Indicators (KEIs) are key to fostering circular systems engineering. 
Capabilities for the analysis and optimization of single lifecycle phases for often contradictory functional and sustainability goals are required that also translate the rather global and socio-economical metrics like SDG targets and even KPIs toward sustainability-related metrics like KEIs~\cite{oecd2008kei}.
As a foundation for this trade-off analysis and optimization, methods are required that foster the translation of the generic metrics to fit the context of the current organization and domain.
\subsection{Digital enablers}\label{sec:challenges-digital}

The reliance on digital enablers in maturation toward circular systems engineering (\secref{sec:levels-strategies}) necessitates advanced engineering techniques of digital technology.

Digital twins are well-positioned to govern end-to-end processes~\cite{heithoff2023digital}, contributing to important operational goals, such as adaptive control and predictive maintenance~\cite{rolnick2022tackling} for sustainability. However, \citet{tzachor2022potential} report challenges that might prevent digital twins from supporting sustainability goals. The level of digital maturity (or the lack thereof) often hinders proper data management, sometimes to the point where real-time data may not be available or is of poor quality. According to McKinsey's industry digitalization index~\cite{hbr2016industrydigitalindex}, particularly challenged are agriculture, construction, and healthcare---three sectors strongly linked to sustainable development through precision agriculture, smart cities, and more patient-oriented health systems, respectively. The feasibility of adopting digital twins is also far from being trivial~\cite{david2023digital}, and it is usually a challenge outside of well-developed countries. Furthermore, modeling and simulation of systems with social elements is a wicked problem~\cite{coyne2005wicked}, and studies on applying digital twins to social problems are currently lacking.
To tackle the challenges of siloed environments and leverage the benefits of digital twinning, interoperability mechanisms of digital twins are to be developed~\cite{piroumian2021digital} to allow for loosely coupled digital twins to scale up into robust hierarchies.

\section{Conclusion}\label{sec:conclusion}

As sustainability is becoming a first-class citizen in modern systems, our systems engineering methods need to be revised to support sustainability ambitions. In this paper, we have defined circular systems engineering, a novel paradigm for systems and method sustainability. Circular systems engineering is the paradigm of designing, developing, operating, maintaining, and retiring systems through sustainable systems principles that foster value retention over multiple engineering cycles.

While the idea of circularity has been around for decades and is recognized as a desirable direction for humankind's overall sustainability and innovation endeavors, systems engineering has yet to adopt sustainability as a governing principle. With the looming paradigm shift from a technology-focused Industry 4.0 towards a value-focused Industry 5.0, systems engineering now has a rare opportunity to promote sustainability to a first-class citizen.
We believe circular systems engineering can be a guiding principle along this road.

To drive the adoption of circularity principles and instigate future research, we have derived prerequisites for circular systems engineering, proposed a sustainability maturity model together with maturity maturation strategies, and identified key challenges and opportunities.
Researchers can use the paradigm to contextualize their sustainability-themed research. Practitioners can rely on the presented maturity framework and maturation strategies to identify the modeling, simulation, and digital transformation goals appropriate to their organizations' sustainability ambitions.

\bibliographystyle{spbasic}
\bibliography{bib/references}


\end{document}